\documentclass[twocolumn,prl,showpacs,amsmath,amssymb,nofootinbib,superscriptaddress,preprintnumbers]{revtex4}

\usepackage{dcolumn}
\usepackage{amsmath}
\usepackage{graphicx}
\usepackage{rotating}
\usepackage{amsfonts}
\usepackage{amssymb}
\usepackage[hang]{subfigure}

\usepackage{graphicx}
\usepackage{psfig,color}
\usepackage{epsfig}

\begin{document}

\preprint{DESY 06-105}

\title {Polarized Light Propagating in a Magnetic Field as a Probe of Millicharged Fermions}

\author{Holger Gies}
\email{h.gies@thphys.uni-heidelberg.de}
\affiliation{Institut f\"ur Theoretische Physik, Universit\"at Heidelberg, Philosophenweg 16, 
D-69120 Heidelberg, Germany}
\author{Joerg Jaeckel}
\email{jjaeckel@mail.desy.de}
\affiliation{Deutsches Elektronen-Synchrotron DESY, Notkestra\ss e 85, D-22607 Hamburg, Germany} 
\author{Andreas Ringwald}
\email{andreas.ringwald@desy.de}
\affiliation{Deutsches Elektronen-Synchrotron DESY, Notkestra\ss e 85, D-22607 Hamburg, Germany}

\begin{abstract}
Possible extensions of the standard model of elementary particle physics suggest
the existence of particles with small, unquantized electric charge. 
Photon initiated pair production of millicharged fermions in an external magnetic
field would manifest itself as a vacuum magnetic dichroism. 
We show that laser polarization experiments searching for this effect yield, in the
mass range below 0.1 eV, much stronger constraints 
on millicharged fermions than previously considered laboratory searches. Vacuum magnetic 
birefringence originating from virtual pair production gives a slightly better constraint for masses between
0.1 eV and a few eV.
We comment on the possibility that the vacuum magnetic dichroism observed by
PVLAS arises from pair production of such millicharged fermions rather than 
from single production of axion-like particles. Such a scenario can be 
confirmed or firmly excluded by a search for invisible decays of orthopositronium 
with a sensitivity of about 
$10^{-9}$  in the corresponding branching fraction.      
\end{abstract}


\maketitle

The apparent quantization of the electric charges of all known elementary particles 
-- i.e., the fact that they appear to be integer multiples
of the electric charge of the $d$ quark -- is a long standing puzzle of fundamental interest.
Strong experimental upper limits on the electric charge of neutrons, atoms, and 
molecules~\cite{Marinelli:1983nd,Dylla:1973,Baumann:1988ue}, $Q < {\mathcal O}(10^{-21})\,e$,
with the magnitude of the electron electric charge  $e$, 
as well as on the magnetic moments of the neutrinos~\cite{Kyuldjiev:1984kz}, 
$\mu_{\nu}< {\mathcal O}(10^{-10})\,\mu_B$, with the Bohr magneton $\mu_B=e/2m_e$ and 
 the electron mass $m_e$, 
strongly support the idea that charge quantization is a fundamental principle. However, the standard
model of particle physics with three generations of quarks and leptons does not impose charge 
quantization~\cite{Foot:1990mn}. One needs physics beyond the standard model in order
to enforce it, as is demonstrated by Dirac's seminal argument for charge
quantization based on the hypothetical existence of magnetic
monopoles~\cite{Dirac:1931kp}. Whereas some extensions of the standard model, e.g. 
 grand unified theories, 
provide mechanisms for enforcing charge quantization, other possible extensions 
suggest the existence of particles of small, unquantized charge $Q_\epsilon = \epsilon e$, 
with $\epsilon\ll 1$~\cite{Ignatiev:1978xj,Okun:1983vw,Holdom:1985ag,Abel:2003ue,Batell:2005wa}.   
 
\begin{figure}[b]
\begin{center}
\includegraphics[bbllx=51,bblly=225,bburx=575,bbury=608,width=8.5cm]{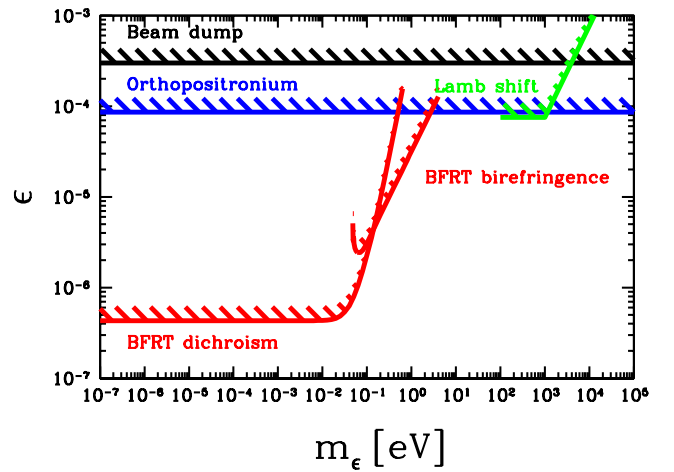}
\end{center}
\caption[...]{Laboratory-based upper limits on the fractional electric charge 
$\epsilon = Q_\epsilon/e$ of a hypothetical millicharged fermion of mass $m_\epsilon$.
The ``Beam dump'' limit has been derived in Ref.~\cite{Davidson:1991si} from a 
beam-dump search for new neutrino-like particles at SLAC~\cite{Donnelly:1978ty,Rothenberg:1982}. 
The ``Orthopositronium'' limit stems from a limit on the branching
fraction of invisible orthopositronium decay~\cite{Mitsui:1993ha}. 
The ``Lamb shift'' limit comes from a recent comparison~\cite{Davidson:2000hf} 
of Lamb shift measurements~\cite{Lundeen:1981,Hagley:1994} 
with QED predictions. 
The ``BFRT dichroism/birefringence'' limit arises from the upper limit on 
vacuum magnetic dichroism/birefringence placed by the laser polarization experiment 
BFRT~\cite{Cameron:1993mr} (see text). 
\label{fig:lablimits}} 
\end{figure}

There are a number of experimental and observational bounds on the fractional electric charge
$\epsilon$ and on the mass $m_\epsilon$ of hypothetical millicharged particles, coming from
laboratory experiments, astrophysics and cosmology~\cite{Dobroliubov:1989mr,Davidson:1991si,%
Mohapatra:1990vq,Mohapatra:1991as,Davidson:1993sj,Prinz:1998ua,Dubovsky:2003yn} 
(for a recent review and further references, see Ref.~\cite{Davidson:2000hf}). 
In the sub-electron mass region, $m_\epsilon < m_e$, the best
laboratory-based 
bounds on millicharged fermions, $\epsilon<{\mathcal O}(10^{-4})$, 
come from searches for the invisible decay of orthopositronium~\cite{Mitsui:1993ha}
and from a comparison~\cite{Davidson:2000hf} of Lamb-shift measurements~\cite{Lundeen:1981,Hagley:1994} 
with predictions of quantum electrodynamics (QED) (cf. Fig.~\ref{fig:lablimits}).
Stronger, albeit more model-dependent bounds arise through astrophysical and cosmological
considerations. For example, stellar evolution constraints~\cite{Raffelt:1996} 
yield a bound
$\epsilon < {\mathcal O}(10^{-14})$, for 
$m_\epsilon < {\mathcal O}(10\ {\rm keV})$,
whereas successful big bang nucleosynthesis leads to the restriction $\epsilon < {\mathcal O}(10^{-9})$, 
for $m_\epsilon < {\mathcal O}(1\ {\rm MeV})$.   

In the present Letter, we want to investigate whether searches exploiting laser polarization 
experiments can give competitive constraints on millicharged fermions, most notably in comparison 
to other laboratory searches.    

It is theoretically well established in QED that photon-initiated electron-positron pair production, 
$\gamma\to e^+ e^-$, in an external magnetic field~\cite{Robl:1952,Toll:1952,Klepikov:1954,%
Erber:1966vv,Baier:1967,Klein:1968,Adler:1971wn,Tsai:1974fa,Daugherty:1984tr,Dittrich:2000zu} 
manifests itself
as a vacuum magnetic dichroism: the polarization vector 
of an initially linearly polarized photon beam with energy 
$\omega > 2m_e$  in general is rotated after passing
a transverse magnetic field. However, because of its high
threshold energy, this effect has not been detected in the
laboratory yet. 
Recent past, present day, and near future instruments for the detection 
of vacuum magnetic birefringence and dichroism, such as BFRT~\cite{Cameron:1993mr}, 
PVLAS~\cite{Zavattini:2005tm}, Q\,\&\,A~\cite{Chen:2003tp}, 
BMV~\cite{BMV}, and proposed experiments at CERN~\cite{Duvillaret:2005sv} and
in Jena~\cite{Heinzl:2006xc}  
exploit photon beams with energies $\omega = {\mathcal O}({\rm eV})$. 
Correspondingly, they may be sensitive to vacuum magnetic dichroism 
induced by the production of fermion anti-fermion pairs with mass 
$2 m_\epsilon < \omega = {\mathcal O}({\rm eV})$. 
Similarly, they may also be sensitive to vacuum magnetic birefringence caused
by the virtual production of these light millicharged particles, which
induces ellipticity of the laser beam in the magnetic field.

Let us first consider dichroism.
Let $\vec k$ be the momentum of the incoming photon, with $|\vec k|=\omega$, and let $\vec B$ 
be a static homogeneous magnetic field, which is perpendicular to $\vec k$, as it
is the case in all of the above-mentioned polarization experiments. The photon-initiated  
production of a Dirac-type fermion anti-fermion pair, with electric charge $Q_\epsilon = \epsilon e$ 
and mass $m_\epsilon$, at $\omega > 2 m_\epsilon$, leads to a non-zero difference between the photon 
absorption coefficients $\kappa_\parallel$ and $\kappa_\perp$,
corresponding to photon polarizations parallel or perpendicular to $\vec B$. 
The fact that the absorption coefficients for the two polarizations, $\parallel$ and $\perp$,
are different directly leads to dichroism:  for a linearly polarized photon beam, 
the angle $\theta$ between the initial polarization vector and the magnetic field will change  
to $\theta + \Delta \theta$ after passing a distance $\ell$ through the 
magnetic field, with  
\begin{equation}
\cot (\theta+\Delta\theta)=\frac{E_{\parallel}}{E_{\perp}}
=\frac{E^{0}_{\parallel}}{E^{0}_{\perp}}
\exp\left(-\frac{1}{2}(\kappa_{\parallel}-\kappa_{\perp})\ell\right).
\label{eq1}
\end{equation}
Here, $E_{\parallel,\perp}$ are the electric field components of the
laser parallel and perpendicular to the external
magnetic field and the superscript ``0'' denotes initial values. 
For small rotation angle $\Delta\theta$, we have  
\begin{equation}
\Delta\theta \simeq \frac{1}{4}(\kappa_{\parallel}-\kappa_{\perp})\ell\, \sin(2\theta).
\end{equation}
Explicit expressions for the photon absorption coefficients $\kappa_{\parallel,\perp}$ can
be inferred from the literature on $\gamma\to e^+e^-$ in a homogeneous magnetic 
field~\cite{Toll:1952,Klepikov:1954,Erber:1966vv,Baier:1967,Klein:1968,Adler:1971wn,Tsai:1974fa,%
Daugherty:1984tr,Dittrich:2000zu}: 
\begin{eqnarray}
\kappa_{\parallel,\perp}\ell 
&=& \frac{1}{2}\epsilon^3 e \alpha \frac{B \ell }{m_\epsilon}\,
T_{\parallel,\perp}(\chi )
\\[1.5ex]
\nonumber
&=& 1.09\times 10^6\ \epsilon^3 
\left( \frac{\rm eV}{m_\epsilon} \right)
\left( \frac{B}{\rm T}\right)
\left( \frac{\ell}{\rm m}\right)\,T_{\parallel,\perp}(\chi )
,
\end{eqnarray} 
where $\alpha = e^2/4\pi$ is the fine-structure constant. 
Here, $T_{\parallel,\perp }(\chi )$ has the form of a parametric 
integral~\cite{Tsai:1974fa},   
\begin{eqnarray}
\label{absorb}
\nonumber
T_{\parallel,\perp} &= & 
\frac{4\sqrt{3}}{\pi\chi}
\int\limits_0^1 {\rm d}v\  
K_{2/3}\left( \frac{4}{\chi}\frac{1}{1-v^2}\right)
\\ 
&& \mbox{} \hspace{10ex} \times 
\frac{\left[ \left( 1-\frac{1}{3}v^2\right)_\parallel,
\left(\frac{1}{2} +\frac{1}{6}v^2\right)_\perp 
\right]}{(1-v^2)}
\\[2ex]
\nonumber
& = & 
\left\{
\begin{array}{clc}
\sqrt{\frac{3}{2}}\ {\rm e}^{-4/\chi}\ \left[(\frac{1}{2})_\parallel,(\frac{1}{4})_\perp\right] & {\rm for} & 
\chi\ll 1 \\
\frac{2\pi}{\Gamma\left(\frac{1}{6}\right)\Gamma\left(\frac{13}{6}\right)} 
\chi^{-1/3}\left[ (1)_\parallel,(\frac{2}{3})_\perp\right]& {\rm for} & \chi\gg 1
\end{array}
\right.
,
\end{eqnarray}
the dimensionless parameter $\chi$ being defined as 
\begin{equation}
\chi \equiv  \frac{3}{2} \frac{\omega}{m_\epsilon} \frac{\epsilon e B}{m_\epsilon^2}
= 88.6\ \epsilon\ \frac{\omega}{m_\epsilon}\  
\left( \frac{\rm eV}{m_\epsilon}\right)^2
\left( \frac{B}{\rm T}\right) 
\,.
\end{equation}
The above expression has been derived in leading order in an expansion
for high frequency, 
\begin{equation}
\label{semiclhf}
\frac{\omega}{2m_\epsilon}\gg  1,
\end{equation}
and of high number of allowed Landau levels of the millicharged particles, 
\begin{subequations}
\begin{equation}
\label{semiclwf}
N_{\rm{Landau}}=\frac{1}{24}\left(\frac{\omega^{2}}{\epsilon\, eB}\right)^{2}\gg 1.
\end{equation}

Let us remark that expression \eqref{absorb} was originally derived in Ref.~\cite{Tsai:1974fa} 
in the more restrictive high-frequency $\omega/2m_\epsilon\gg 1$ and weak-field limit
$\epsilon\,eB/m_\epsilon^2\ll 1$, in agreement with the results of 
Refs.~\cite{Toll:1952,Klepikov:1954,Erber:1966vv,Baier:1967,Klein:1968}.
These in turn agree with the result of 
Ref.~\cite{Daugherty:1984tr} which are obtained with the conditions 
\eqref{semiclhf} and \eqref{semiclwf}. 
Intuitively, we can understand the nature of this approximation as follows. 
Expression \eqref{absorb} is a rather smooth function of the frequency $\omega$. However, from the discrete 
nature of the Landau levels we would rather expect absorption peaks. Yet, if the peaks are very 
dense we cannot resolve them and we have to average over a small frequency interval 
$\Delta\omega$, yielding the smooth function \eqref{absorb}. Averaging is allowed if we have a 
large number of peaks $\Delta N_{\rm{peaks}}$ in the interval $\Delta\omega$,
\begin{eqnarray}
\label{peaks}
\Delta N_{\rm{peaks}}\!\!&=&\!\!\frac{1}{2}\Delta N_{\rm{Landau}}=
\frac{1}{12}\left(\frac{\omega^{2}}{\epsilon\,eB}\right)^{2}\frac{\Delta\omega}{\omega}\gg 1
\\[1.5ex]\nonumber
&&\Leftrightarrow \epsilon\ll 4.89\times 10^{-3} \left(\frac{\omega}{\rm{eV}}\right)^{2}
\left(\frac{\rm{T}}{B}\right)
\left(\frac{\Delta\omega}{\omega}\right)^{\frac{1}{2}}.
\end{eqnarray}
\end{subequations}
This expression agrees with \eqref{semiclwf} up to a factor of $\Delta\omega/(2\omega)$ which takes 
the uncertainty in the frequency into account. In the above-mentioned laser polarization experiments,
a cavity is used to reflect the laser beam back and forth, thereby enhancing the signal. 
In that case, the frequency uncertainty is $\Delta\omega/\omega\sim 1/N_r$, where $N_r$ is the 
number of reflections in the cavity.  

At present, the most stringent bound on vacuum magnetic dichroism comes 
from the BFRT laser polarization experiment~\cite{Cameron:1993mr}. 
A linearly polarized laser beam ($\omega = 2.41$~eV) was sent along the magnetic field of two superconducting 
dipole magnets ($B=2$~T), which were placed in an optical cavity with $N_r=254$ reflections, such that
the optical path length was $\ell = N_r\times 8.8\ {\rm m}\simeq 2235$~m. 
An upper limit on the absolute value of the rotation,  
\begin{equation}
|\Delta\theta| < 6\times 10^{-10} \ 
\hspace{4ex} (95\,\%\ {\rm confidence\ level})
, 
\end{equation}
was obtained. This can be turned into an upper limit on $\epsilon$, as a function
of $m_\epsilon$, by exploiting the predictions
\eqref{eq1}-\eqref{absorb} for $\Delta\theta$ 
from photon-initiated pair production of millicharged fermions in an external
magnetic field. The resulting limit is displayed in Fig.~\ref{fig:lablimits}
and labelled as ``BFRT dichroism".  
Clearly, for small masses, $m_\epsilon \lesssim 0.1$~eV, this 
represents currently the best laboratory limit on millicharged fermions. 

Let us now turn to birefringence. The propagation speed of the laser
photons is slightly changed in the magnetic field owing to the
coupling to virtual charged pairs.
The corresponding refractive indices $n_{\parallel,\perp}$ differ for
the two polarization modes, causing a phase difference between the
two modes,
\begin{equation}
\Delta\phi=\omega \ell(n_{\parallel}-n_{\perp}). 
\end{equation}
This induces
an ellipticity $\psi$ of the outgoing beam,
\begin{equation}
|\psi|=\frac{\omega\ell}{2}|(n_{\parallel}-n_{\perp})\sin(2\theta)|\quad\quad\rm{for}\,\,\psi\ll1.
\end{equation}

Virtual production can occur even below threshold $\omega<2m_{\epsilon}$. Therefore, 
we consider both high and low frequencies.
As long as \eqref{peaks} is satisfied one has~\cite{Tsai:1975iz}
\begin{equation}
n_{\parallel,\perp}=1-\frac{\epsilon^{2}\alpha}{4\pi}\left(\frac{\epsilon\,eB}{m^{2}_{\epsilon}}\right)^{2}
I_{\parallel,\perp}(\chi),
\end{equation}
with
\begin{eqnarray}
\nonumber
I_{\parallel,\perp}(\chi)\!\!&=&\!\!2^{\frac{1}{3}}\left(\frac{3}{\chi}\right)^{\frac{4}{3}}
\int^{1}_{0} {\rm d}v\,
\frac{\left[\left(1-\frac{v^2}{3}\right)_{\parallel},
\left(\frac{1}{2}+\frac{v^2}{6}\right)_{\perp}\right]}{(1-v^{2})^{\frac{1}{3}}}
\\[1.5ex]
&&\quad\quad\quad\quad\quad\quad\quad\quad\quad\quad\times
\tilde{e}^{\prime}_{0}\left[\begin{scriptstyle}-
\left(\frac{6}{\chi}\frac{1}{1-v^2}\right)^{\frac{2}{3}}\end{scriptstyle}\right]
\label{refrac}\\[1.5ex]\nonumber
&&\!\!\!\!\!\!\!\!\!\!\!\!\!\!\! =  
\!\!\left\{
\begin{array}{clc}  - 
\frac{1}{45} \left[(14)_\parallel,(8)_\perp\right] & {\rm for} & \chi\ll 1 \\
\frac{9}{7}\frac{\pi^{\frac{1}{2}}2^{\frac{1}{3}}
\left(\Gamma(\left(\frac{2}{3}\right)\right)^{2}}{\Gamma\left(\frac{1}{6}\right)} 
\chi^{-4/3}\left[ (3)_\parallel,(2)_\perp\right]& {\rm for} & \chi\gg 1
\end{array}
\right.
.
\end{eqnarray}
Here, $\tilde{e}_{0}$ is the generalized Airy function,
\begin{equation}
\tilde{e}_{0}(t)=\int^{\infty}_{0}{\rm d}x\,\sin\left(tx-\frac{x^3}{3}\right),
\end{equation}
and 
$\tilde{e}^{\prime}_{0}(t)={\rm{d}}\tilde{e}_{0}(t)/{\rm{d}}t$.
Using the parameters for the BFRT birefringence measurement, $\omega =2.41$~eV, $B=2$~T, 
$N_r=34$, and $\ell = N_r\times 8.8$~m,  their upper limit on the ellipticity,
\begin{equation}
|\psi|<2\times10^{-9}\quad (95\%\ \rm{confidence\ level}), 
\end{equation}
leads to  the limit depicted in Fig.~\ref{fig:lablimits}, 
  which is the
currently best laboratory limit in the range $0.1\ \text{eV} \lesssim m_\epsilon
\lesssim 3\ \text{eV}$. 
Let us finally remark that all our limits remain
valid for $m_\epsilon \gtrsim 10^{-2}$~eV, even if we impose the more strict validity
constraint $\epsilon eB/m_\epsilon^2\ll 1$ for Eqs.~\eqref{absorb} and
\eqref{refrac}. For a check of the quantitative convergence of the
underlying expansion for $m_\epsilon \lesssim 10^{-2}$~eV, a next-to-leading order
calculation may ultimately be needed.

\begin{figure}
\begin{center}
\includegraphics[bbllx=51,bblly=225,bburx=575,bbury=608,width=8.5cm]{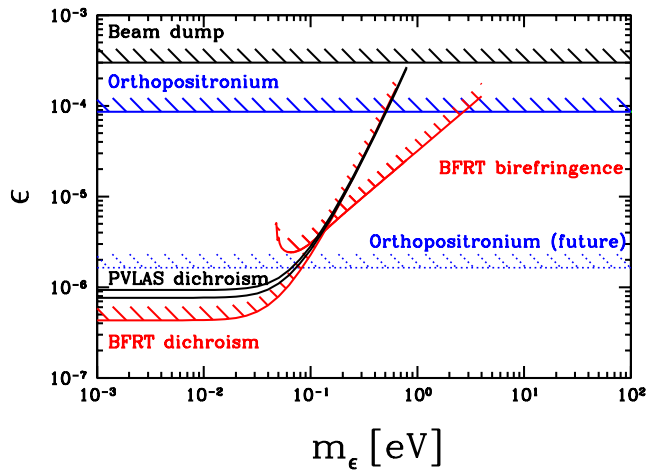}
\end{center}
\caption[...]{Laboratory-based upper limits on the fractional electric charge 
$\epsilon = Q_\epsilon/e$ of a hypothetical millicharged fermion of mass $m_\epsilon$ 
(same as in Fig.~\ref{fig:lablimits}). 
The parameter values between the two lines labelled ``PVLAS dichroism'' 
correspond to the preferred 95\,\% confidence region if the PVLAS rotation is 
interpreted as orginating
from pair production of millicharged fermions. 
The dashed limit labelled  ``Orthopositronium (future)"  
corresponds to the projected 95\,\% exclusion limit obtainable through a 
search for invisible orthopositronium decay with a 
sensitivity of $10^{-9}$ in the corresponding branching ratio. 
\label{fig:pvlas}} 
\end{figure}

Recently, the PVLAS collaboration reported the observation of an optical rotation 
generated in vacuum by a magnetic field~\cite{Zavattini:2005tm},
\begin{equation}
|\Delta\theta|/N_r = (3.9\pm 0.5)\times 10^{-12} 
.
\end{equation}
The experimental parameters in their setup were $\omega =1.17$~eV, $B=5$~T, 
$N_r=4.4\times 10^4$, and $\ell = N_r\times 1$~m. 
If interpreted in terms of pair production of millicharged fermions, we 
obtain the preferred 95\,\% confidence region lying between the two black lines labelled
``PVLAS dichroism'' in Fig.~\ref{fig:pvlas}. 
Apparently, at two standard deviations, this is in conflict with the
limit from BFRT.  
Nevertheless, the PVLAS result is very close to the boundary
of the excluded region for masses ${\mathcal{O}(0.1\,\rm{eV}})$, 
and therefore the pair-production interpretation still 
represents a remote   
alternative to the standard, axion-like-particle (ALP) interpretation of the 
PVLAS result~\cite{Maiani:1986md,Ringwald:2005gf}. For both interpretations, there are 
problems with the 
astrophysical bounds~\cite{Raffelt:1996} which are difficult to 
avoid for ALPs~\cite{Masso:2005ym,Jain:2005nh,Jaeckel:2006id,Masso:2006gc}. 
It remains to be seen whether pair production provides easier ways to evade them. 
A promising way to test the parameter region around $m_{\epsilon}\sim 0.1\ \rm{eV}$, 
$\epsilon \sim 3\times 10^{-6}$,   
opens up in the near future, when the sensitivity of 
searches for the invisible decay of orthopositronium reach the 
$10^{-9}$
level in the corresponding branching ratio~\cite{Rubbia:2004ix} 
(cf. Fig.~\ref{fig:pvlas}). Also, a PVLAS birefringence
measurement can be expected to explore the interesting region around
$m_{\epsilon}\sim 0.1\ \rm{eV}$; a positive signal would fix both
parameters $\epsilon$ and $m_\epsilon$ of hypothetical millicharged
particles by reading off the intersection point of the dichroism and
birefringence curves. 

{\it In summary,} polarization measurements of laser beams traversing intense magnetic
fields provide a very sensitive probe of light millicharged fermions in the 
laboratory. In the sub-eV range, already the limits inferred from the pioneering BFRT experiment 
are more than two orders of magnitude better than 
other laboratory based limits.

\vspace*{0.5cm}
\centerline{\bf Acknowledgments}
\vspace{0.05cm}
A.R. would like to thank Eduard Masso, Javier Redondo, 
Carlo Rizzo, and Giuseppe Ruoso for an inspiring discussion 
on electric charge quantization and neutrality of neutrons and atoms. 
HG acknowledges support by DFG Gi 328/1-3.

\end{document}